\documentclass[aip,reprint
,amsmath,amssymb
]{revtex4-1}

\usepackage{graphicx}
\usepackage{xcolor}
\usepackage{dcolumn}
\usepackage{bm}
\usepackage{soul}

\usepackage[utf8]{inputenc}
\usepackage[T1]{fontenc}
\usepackage{mathptmx}
\usepackage{etoolbox,hyperref}

\draft 

\begin{document}

\preprint{AIP/123-QED}

\title[Negativity Bias]{Decision-Making under Negativity Bias: Double Hysteresis in the Opinion-Dependent $q$-Voter Model}

\author{Maciej Doniec}
\email{maciej.doniec@pwr.edu.pl}
\affiliation{Department of Computational
Social Science, Wrocław University of Science and Technology, Poland}

\author{Katarzyna Sznajd-Weron}
\email{katarzyna.weron@pwr.edu.pl}
\affiliation{Department of Computational Social Science, Wrocław University of Science and Technology, Poland}

\author{Federico Vazquez}
\email{fede.vazmin@gmail.com}
 \homepage{
\\
https://fedevazmin.wordpress.com}
\affiliation{Instituto de Cálculo, FCEN, Universidad de Buenos Aires and CONICET, C1428EGA Buenos Aires, Argentina} \affiliation{Institute for Cross-disciplinary Physics and Complex Systems IFISC (CSIC-UIB), Campus Universitat Illes Balears, 07122 Palma de Mallorca, Spain}

\date{\today}

\begin{abstract}

Negative information often exerts a disproportionately strong impact on human decision-making, a phenomenon known as the negativity bias. In behavioral economics, this effect is formally captured by Prospect Theory, which posits that losses loom larger than equivalent gains. For example, a single negative product review can outweigh numerous positive ones, reflecting this principle of loss aversion in consumer behavior. While this psychological effect has been widely documented, its implications for collective opinion dynamics, critical for understanding market stability and reputation dynamics, remain poorly understood. Here, we generalize the $q$-voter model with independence by introducing opinion-dependent influence group sizes, $q_+$ and $q_-$, which represent the social reinforcement needed to change an opinion from negative to positive and from positive to negative, respectively. We study two versions of this asymmetric model: a baseline model that reduces to the standard $q$-voter model when $q_+ = q_- = q$, and an extended model that incorporates an additional asymmetry expressed as a preference for one opinion. In its reduced version, this represents a minimal model in terms of non-linearity within the $q$-voter framework that allows for discontinuous phase transitions and hysteresis. Using mean-field analysis and computer simulations, we show that these modifications lead to rich collective behaviors, including double hysteresis, one form of which is irreversible, providing a mechanism for path-dependence and the sustained, irrecoverable damage to collective sentiment, brand equity, or market confidence.
\end{abstract}

\maketitle

\begin{quotation}

Loss aversion, a central principle of Prospect Theory in Behavioral Economics, dictates that losses loom larger than equivalent gains. In market settings, this translates to the negativity bias: a single negative comment can outweigh a dozen positive ones, potentially causing a product or firm's reputation to collapse after just one bad review. While this individual-level asymmetry in judgment strongly influences consumer behavior, its implications for collective opinion dynamics, critical for market sentiment and reputation management, remain largely unexplored. To address this, we extend the $q$-voter model with independence by allowing the size of the influence group to depend on the direction of opinion change: an agent requires a larger group of supporters to switch from negative to positive than from positive to negative. This simple asymmetry captures the essence of negativity bias and significantly alters the system’s macroscopic behavior. Using analytical calculations and simulations, we show that the system can exhibit double hysteresis, combining reversible and irreversible opinion shifts. Furthermore, even very small influence groups can trigger discontinuous phase transitions and hysteresis, features that, in the standard $q$-voter model with independence, appear only for larger group sizes $q>5$. These results identify a minimal mechanism by which individual-level asymmetry in social influence can generate macroscopic irreversibility in collective opinion dynamics.
\end{quotation}

\section{Introduction}
Imagine the following situation: you plan to buy new headphones. After reading numerous positive reviews online, you decide to purchase a particular model. Then, you come across a single negative review, and you cancel the purchase. This scenario illustrates the negativity bias, a psychological tendency for negative information, reviews, or opinions to have a stronger impact than positive ones on how people perceive, share, and act on information \cite{Lei2023PositiveReviews,Yin2016ResearchMouth,Varga2024TheDecisions}. Even when a consumer is exposed to many positive signals (e.g., high ratings, multiple good reviews), a single negative review can outweigh them all and alter the decision, for instance, by causing the person to abandon a purchase. Indeed, recent studies show that negative reviews significantly reduce a product’s purchase probability \cite{Varga2024TheDecisions}.

Negativity bias is closely linked to Prospect Theory, a core concept in behavioral economics \cite{kahneman1979prospect}. According to this theory, people evaluate outcomes relative to a reference point and feel losses more strongly than gains of the same size, a phenomenon known as loss aversion. Recent studies show that Prospect Theory also explains how online ratings relate to review sentiment: negative changes in ratings have a stronger effect on sentiment than positive ones \cite{Sharma2020}. Beyond financial decisions, this asymmetric response to negative versus positive outcomes reflects a general principle that bad is stronger than good \cite{Baumeister2001BadGood}. Negative feedback and emotions draw more attention and have a greater influence than positive ones \cite{Rozin2001296}. From this perspective, the negativity bias observed in online behavior can be seen as a social and experiential form of loss aversion, a deep-seated tendency to focus more on avoiding negative outcomes than on pursuing positive ones.

The observation that even one negative opinion can outweigh multiple positive ones motivated us to generalize the $q$-voter model  ($q$-VM) \cite{Castellano2009NonlinearModel}. Specifically, we generalize its variant with independence \cite{Nyczka2012PhaseDriving}, which is particularly interesting because its binary form exhibits tricriticality and allows for discontinuous phase transitions. This model and its variations, as well as generalizations such as the Abrams-Strogatz and the  nonlinear noisy voter model \cite{Peralta2018AnalyticalNetworks}, have been studied extensively on various network structures \cite{Vazquez2010AgentTransitions,Jedrzejewski2017PairNetworks,Ramirez2024OrderingModels,Krawiecki2024Q-voterApproximations}, including multilayer and interconnected networks \cite{Alvarez-Zuzek2016InteractingNetworks,Gradowski2020PairNetworks,Chmiel2015PhaseClique}.

The original $q$-VM is symmetrical and homogeneous, meaning that agents are identical, neither opinion is preferred, and each agent is influenced by a group of size $q$, a fixed parameter of the model. Several extensions of the original $q$-voter model have been proposed in which the symmetry between the two opinions ($+1$ and $-1$) is preserved, while asymmetry is introduced through heterogeneity in how agents respond to social influence. Mellor, Mobilia, and Zia developed the heterogeneous $q$-voter model with zealotry, where two subgroups of susceptible agents use different group sizes, $q_1$ and $q_2$, when consulting their neighbors \cite{Mellor2016CharacterizationZealotry,Mellor2017HeterogeneousZealotry}. When $q_1 \neq q_2$, the two subgroups evolve under different dynamical rules, breaking detailed balance and driving the system into a genuine nonequilibrium steady state.  Abramiuk-Szurlej \textit{et al.} \cite{Abramiuk-Szurlej2021DiscontinuousGraphs} proposed another type of heterogeneity in group sizes, where the group size depends on the type of response. In this model, the system is homogeneous, meaning that all agents are identical. However, at each update, they randomly choose how to respond to social influence, either as conformists or anticonformists, and the size of the influence group depends on the chosen response. This modification leads to discontinuous phase transitions, which do not occur if the sizes of the influence groups are the same for conformity and anticonformity \cite{Nyczka2012PhaseDriving}. Yet another approach was proposed by Chmiel \textit{et al.} in the asymmetric $q$-voter model on multiplex networks \cite{Chmiel2020ANetworks}. Here, all agents are identical, but the multiplex structure introduces two layers with different lobby sizes ($q_1$ and $q_2$). This structural asymmetry yields novel phenomena such as successive and hybrid phase transitions.  

In all these approaches, the opinions remain fully symmetric; however, heterogeneity in group size, whether across agent types, types of social response, or network layers, creates asymmetry in the dynamics. The asymmetry between opinions has also been explored and introduced to the $q$-voter model in various ways, including through an external field representing the media or advertisement \cite{Civitarese2021ExternalModels,Azhari2023TheGraph,Fardela2024OpinionModel,Muslim2024MassModel}, or simply as a bias towards one of the two opinions \cite{Byrka2016DifficultyPractices,Mullick2025SocialInfluence,Doniec2025ModelingModel,Abramiuk-Szurlej2025PairModel}, where the probability of adopting one opinion is larger due to some independent parameter rather than group size. However, to the best of our knowledge, no study has introduced asymmetry through opinion-dependent group sizes. Therefore, this paper aims to extend the $q$-VM with independence by incorporating such asymmetry. This approach is significant because it captures the psychological phenomenon of negativity bias, where it takes more positive feedback to change an opinion from bad to good than the reverse, and because it leads to rich collective behaviors, including double hysteresis, one form of which is irreversible and can, for example, explain the damaged reputation of a product or company. To investigate these effects, we perform mean-field analysis and computer simulations, focusing on stationary states and phase transitions.

\section{The model}

We consider a population of $N$ agents, where each can hold one of two possible states, $S=\pm 1$, representing two different behaviors: for instance, to decide to buy a new product ($S=+1$) or not ($S=-1$). Each agent is allowed to switch states when it is influenced by a group of $q$ other agents, chosen at random.  Only if all $q$ agents in the influencing group share the same state $S$, then the agent adopts the state $S$.  In addition, agents can also switch states spontaneously ($S \to -S$).  A parameter $p$ controls the relative frequency of these two processes, as we shall detail below.  

Despite the similarity with the original version of the $q$-VM, here we consider that the size of the influencing group depends on the choice. That is, an individual needs to be influenced by $q_+$ individuals that have chosen to buy the product to decide to buy, and by $q_-$ individuals that have chosen not to buy the product to decide not to buy. As a few negative reviews are enough to reject the purchase, whereas many positive reviews are needed for the purchase, it turns natural to consider $q_- < q_+$. However, to explore all possible scenarios, we leave $q_+$ and $q_-$ as two free independent parameters of the model.  In the particular case that both choices are considered equivalent, we then set $q_+=q_-=q$, and thus the model reduces to the original version of the $q$-VM with independence and conformity. 

\begin{figure*}
\centering
\includegraphics[width=0.8\textwidth]{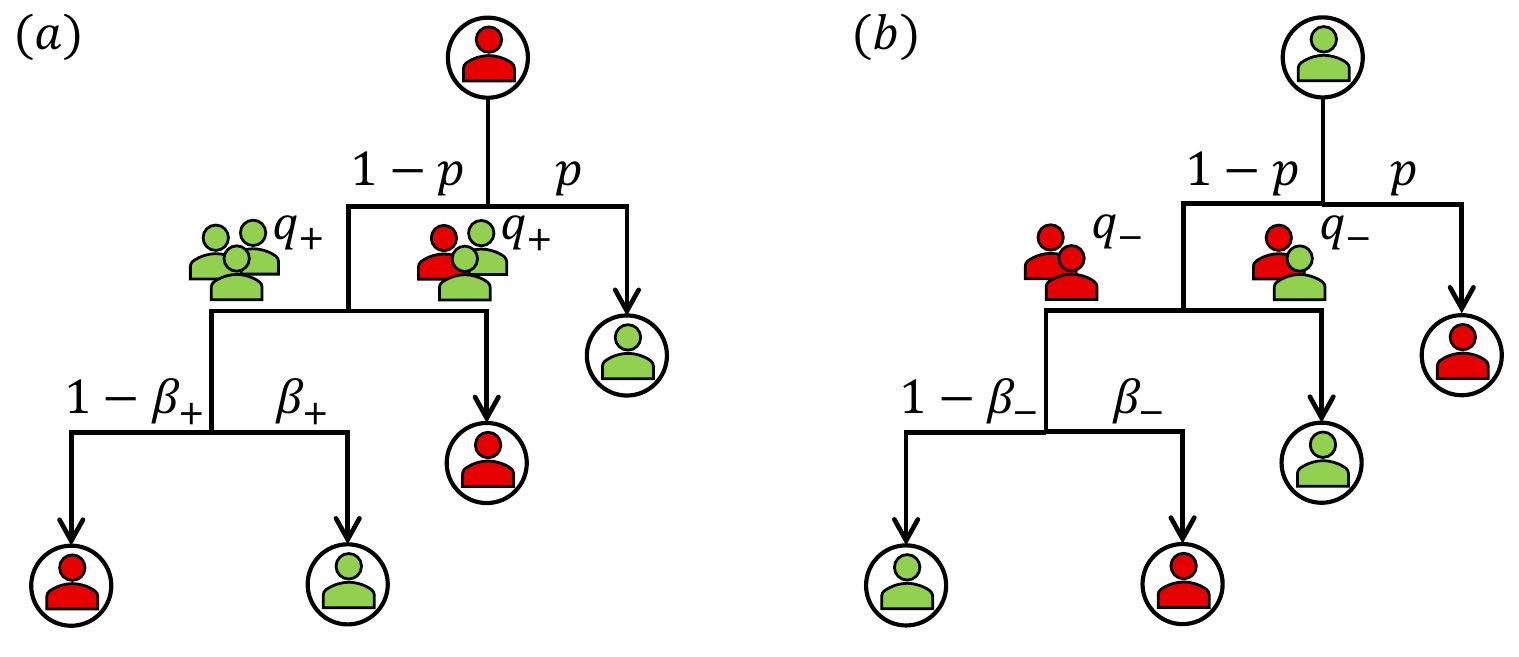}
\caption{Illustration of the extended model, where red denotes negative opinions, green positive opinions, and target agents are shown in circles: (a) A negative agent may, with probability $p$, act independently and switch to positive, or, with probability $1-p$, be influenced by the $q$-panel, but only if it is unanimous; otherwise, it retains its state. When exposed to a unanimous positive $q$-panel, a negative agent conforms with probability $\beta_+$ and resists with the complementary probability. (b) The process is analogous for a positive agent, but the influence group sizes in (a) and (b) differ: $q_+$ is the group size required for a shift from negative to positive, and $q_-$ the size required for the reverse. For the baseline model $\beta_+=\beta_-=1$.}
    \label{fig:diagram_beta}
\end{figure*}

\subsection{Baseline model}

We consider the $q$-VM with independence \cite{Nyczka2012PhaseDriving} and introduce an asymmetry in the interactions by allowing each agent to adopt a given state $S=\pm 1$ after interacting with a group of $q_s$ agents, i.e., an influencing group whose size $q_s$ depends on $S$. At the initial time $t=0$, a fraction $c(0)$ of agents chosen at random is endowed with state $S=+1$, while the complementary fraction $1-c(0)$ takes the state $S=-1$. Then, the dynamics is as follows.  In a single time step $\Delta t=1/N$, an agent with state $S$ is chosen at random.  Then, with probability $p$ the agent flips state (independence) or, with the complementary probability $1-p$, the agent updates its state according to a random group of agents: $q_{-s}$ other agents are randomly chosen, and if all $q_{-s}$ agents are in the same state $-S$ (opposite to the agent's state), then the agent flips its state to match that of the influencing group.  This step is repeated until the system eventually reaches a steady state. It is worth noting that in the $q$-VM model with independence \cite{Nyczka2012PhaseDriving}, a slightly different definition of independence was used, what could be called random resetting noise: with probability  $p$ the agent randomly chooses state $+1$ or $-1$ each with probability $1/2$. In contrast, here we use pure flip noise: with probability $p$ the agent simply flips its state, analogous to the definition used in the nonlinear noisy voter model \cite{Peralta2018AnalyticalNetworks}.

In the case that $q_{+}>q_{-}$, $-1$ agents would be less likely to switch state ($-1 \to +1$), as they require a larger group with an unanimous option $+1$ to change.  In general, we expect that the dynamics favor the option with the largest influence group, as agents with that option would have a larger inertia to keep their present state.     

\subsection{Extended model}

The model described above introduces an asymmetry between the two states through the parameters $q_{\pm}$, which control the size of the influence group.  Here we propose an extension of this model that adds an extra source of asymmetry, in the form of a preference for one of the two options.  In concrete, when all $q_{-}$ ($q_{+}$) agents in a randomly chosen influence group of a $+1$ ($-1$) agent are in the opposite state $-1$ ($+1$), the agent switches to state $-1$ ($+1$) with probability $\beta_{-}$ ($\beta_{+}$), as described in Fig. \ref{fig:diagram_beta}.
This extended model reduces to the baseline model when $\beta_+=\beta_-=1$. By choosing for instance $\beta_{-}>\beta_{+}$, the $+1$ agents would tend to flip more often than the $-1$ agents, introducing a bias in the direction of the $-1$ (preferred) option, since agents would be more likely to adopt the $-1$ than the $+1$ option when their influence groups are unanimously opposite.  This bias can be interpreted as an external field acting in the $-1$ direction.

\section{Results}

We perform the analysis of the models described above by means of a mean--field approach that describes the evolution of the system in the limit of a large population ($N \gg 1$), where finite--size effects and correlations are neglected.  This approach is suitable for populations with homogeneous (all--to--all) interactions, i.e., where each agent interacts with any other agent with the same probability, as in the present models.  Analytical results are then compared with Monte--Carlo (MC) simulations of the dynamics, showing a good agreement for large enough populations, as we shall see.  

\begin{figure*}[t]
\centering
\includegraphics[width=0.8\textwidth]{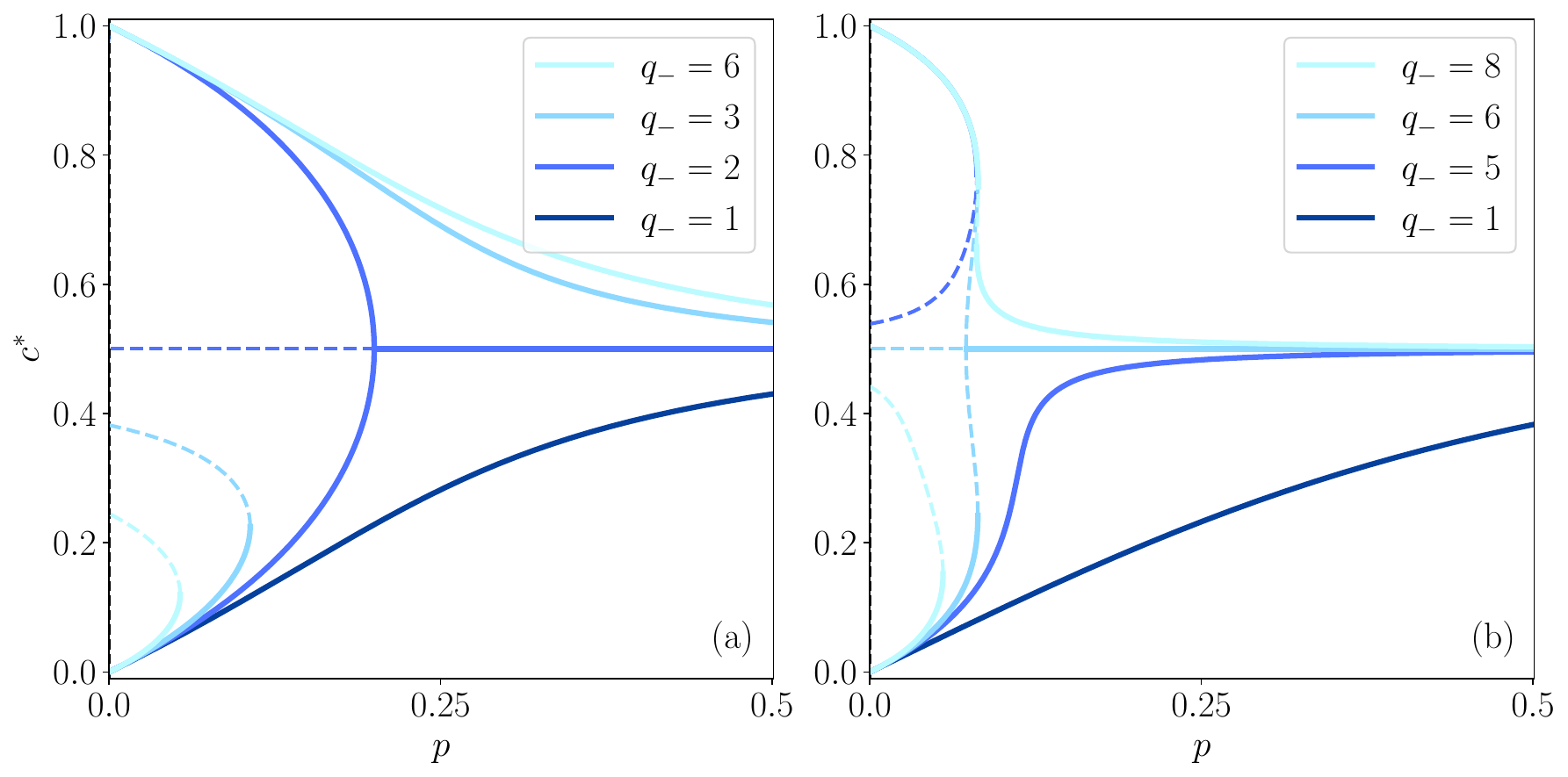}
\caption{\textbf{Baseline model}. The stationary concentration of $+1$ agents $c^*$ as a function of the probability of independence $p$, obtained from Eq.~(\ref{eq:p_stacionary}) for (a) $q_{+}=2$ and (b) $q_{+}=6$, and for the values of $q_-$ indicated in the legends. Solid lines denote stable steady states, while dashed lines correspond to unstable steady states. (a) Three different behaviors are observed: a smooth change of $c^*$ with $p$ for $(q_+,q_-)=(2,1)$ (no transition), a symmetric continuous transition for $(q_+,q_-)=(2,2)$, and an irreversible discontinuous transition for $(q_+,q_-)=(2,3)$ and $(2,6)$ (lower branch). (b) There is a symmetric discontinuous transition for $(q_+,q_-)=(6,6)$ and an irreversible discontinuous transition for $(q_+,q_-)=(6,5)$ (upper branch). The $(q_+,q_-)=(6,8)$ curves show both, a reversible and an irreversible discontinuous transitions at the upper and lower branches, respectively.  The reversible transition is shown in more detail in Fig.~\ref{fig:double_histeresis}.}
    \label{fig:stacionaru_beta1_all_2x2}
\end{figure*}

\subsection{Baseline model}

The macroscopic state of the system at a given time $t$ is well described by the concentration of agents with state $+1$, defined as $c(t)$.  In the $N \gg 1$ limit, the time evolution of $c(t)$ is given by the following rate equation:
\begin{equation}
    \frac{dc}{dt}=(1-p)\left[(1-c)c^{q_{+}} - c(1-c)^{q_{-}} \right] +p(1-2c).
    \label{eq:mfa}
\end{equation}
The first term represents conformity, which occurs with probability $(1-p)$, while the second term represents a random flip of an agent, occurring with probability $p$ due to independence. Thus, the term in square brackets accounts for the change of $c$ after a group interaction. That is, selecting a negative agent with probability $(1-c)$ and flipping its state with the probability $c^{q_{+}}$ that a $q_{+}$--panel of positive agents is randomly chosen (first term), and selecting a positive agent with probability $c$ and flipping its state with the probability $(1-c)^{q_{-}}$ of choosing a $q_{-}$--panel of negative agents (second term). We are interested in the stationary values of $c$ at the steady state, which correspond to the fixed points of Eq.~(\ref{eq:mfa}).  By setting $dc/dt=0$ in Eq.~(\ref{eq:mfa}) and solving for $p$, we obtain the following relation between the fixed points $c^*$ and the independence parameter $p$: 
\begin{equation}
    p= \frac{(1-c^*)c^{*^{q_{+}}}-c^*(1-c^*)^{q_{-}}}{(1-c^*)c^{*^{q_{+}}}-c^*(1-c^*)^{q_{-}}-(1-2c^*)},
    \label{eq:p_stacionary}
\end{equation}
valid for $q_+ \ne 1$ or $q_- \ne 1$.  For the particular case of the linear voter model ($q_+=q_-=1)$ we obtain from Eq.~(\ref{eq:mfa}) the fixed point $c^*=1/2$ for $p>0$ and $c^*=c(0)$ for $p=0$.  In Fig.~\ref{fig:stacionaru_beta1_all_2x2} we plot the fixed points $c^*$ as a function of $p$ from Eq.~(\ref{eq:p_stacionary}) for two different values of $q_+$, $q_+=2$ (a) and $q_+=6$ (b), and four different values of $q_-$ in each panel. Each curve $c^*(p)$ corresponds to a particular set $(q_+,q_-)$. 

In Fig.~\ref{fig:stacionaru_beta1_all_2x2}(a) for $q_{+}=2$ we see that the curve for $q_-=1<q_+$ shows a smooth increase with $p$ that approaches $c^*=1/2$ as $p$ tends to $1.0$, and it is below $1/2$ for all $p<1$. That is, agents with the negative state become dominant in the system. This is because, when $q_-<q_+$, $+1$ agents have a smaller influence group that must be unanimous for them to switch to negative state, and thus it becomes easier for positive agents to switch to negative than vice versa, leading to a larger number of negative agents at the steady state. In general, we observe a negative or a positive dominance depending if $q_-<q_+$ or $q_+<q_-$, respectively. In other words, the option that requires a larger $q$--panel to switch tends to be the dominant option at the steady state, as agents with that option show more resistance or "inertia" to change. 

\begin{figure*}[t]
\centering
\includegraphics[width=0.8\textwidth]{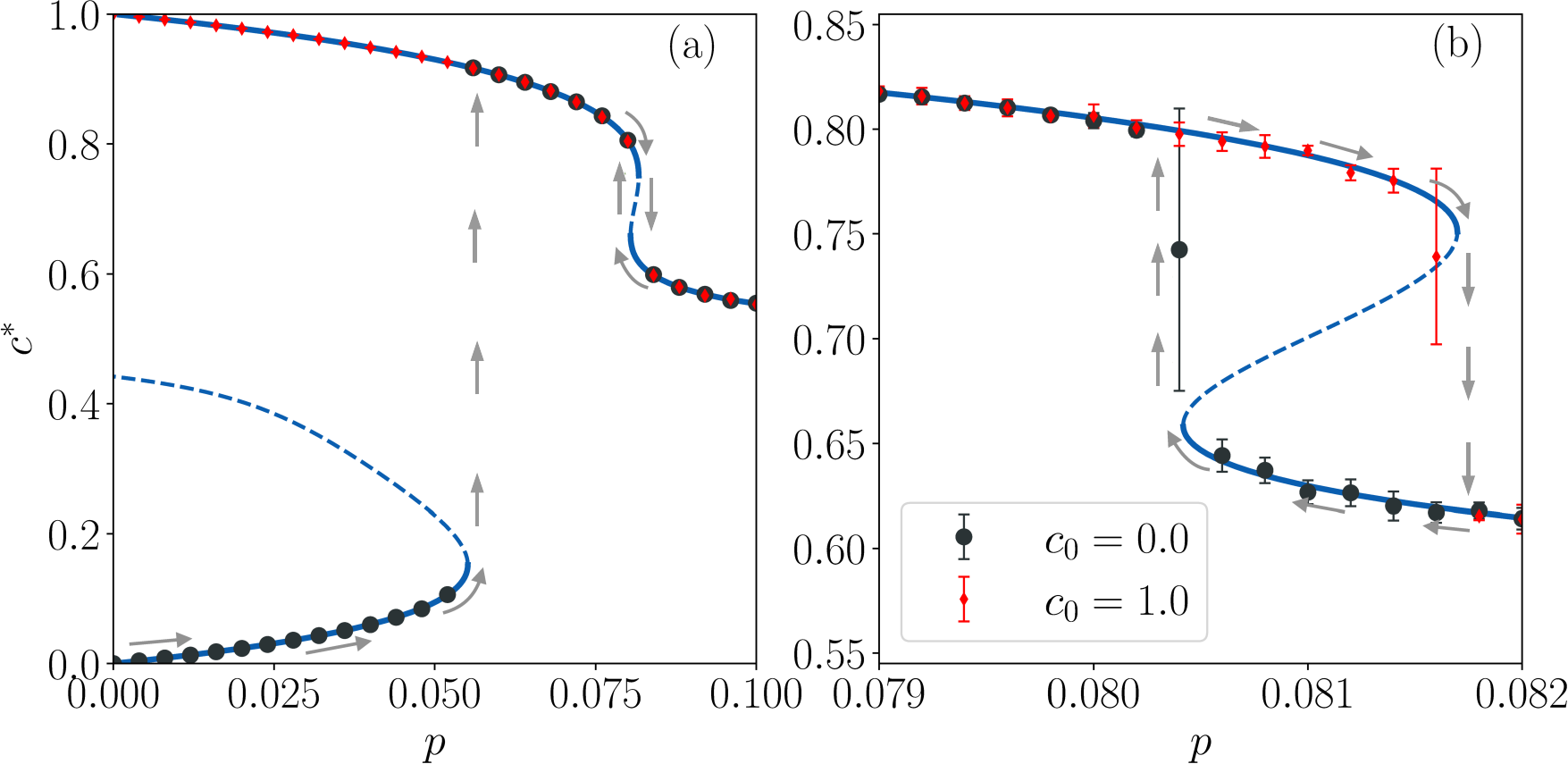}
\caption{\textbf{Baseline model: double hysteresis}. Stationary concentration $c^*$ vs $p$ for $q_{+}=6$ and $q_{-}=8$. Solid and dashed lines correspond to the stable and unstable fixed points from Eq.~\eqref{eq:p_stacionary}, respectively, while symbols are the results from Monte Carlo simulations of the dynamics.  The numerical simulation value of $c^*$ for each $p$ corresponds to the time average of $c(t)$ at the stationary state of a single realization of the dynamics for a system of $N=10^6$ agents, starting from two different initial conditions, $c(0)=0.0$ (circles) and $c(0)=1.0$ (diamonds). (a) Double hysteresis: a reversible hysteresis takes place in the bistable region $0.0805 \lesssim p \lesssim 0.0817$ (upper branch), while an irreversible hysteresis is observed in the bistable region $0 \le p \lesssim 0.055=p_c$ (lower branch).  The former corresponds to a classical hysteresis loop [zoomed in panel (b)], while the later consists of a irreversible hysteresis loop that leads to a cusp catastrophe: once the system jumps from the lower to the upper branch as $p$ overcomes $p_c$, it remains in the upper branch as $p$ is varied.  (b) Zoom in on the region of the reversible hysteresis curve.}  
    \label{fig:double_histeresis}
\end{figure*}

For $q_{+}=q_{-}=2$ we observe that the system recovers the behavior of the symmetric $q$-VM, which displays a continuous transition as $p$ overcomes a threshold value.  However, the transition becomes discontinuous for $q_- > q_+$ ($q_-=3$ and $6$), showing a bistable region in the interval of $0 \le p \le p_c$, defined by an upper and a lower branch that consist of stable fixed points (solid curves), where the transition threshold $p_c$ decreases with $q_{-}$. In the mono-stable region $p>p_c$ we see that $c^*(p)>1/2$ (upper branch), thus positive agents dominate for $q_+<q_-$. However, in the bistable region $0 \le p \le p_c$ both the positive and the negative states can dominate depending on the initial condition, determined by the dashed curve (unstable branch). As in most discontinuous transitions, there is an associated hysteresis curve, but in this case the hysteresis is \emph{irreversible}, because the loop is not complete.  At $p_c$ there is an imperfect pitchfork bifurcation: the upper branch consists entirely of stable fixed points, while the lower branch shows a saddle--node bifurcation at $p_c$, where the stable (solid curve) and unstable (dashed curve) branches meet and disappear.  This half hysteresis loop is associated with a "cusp catastrophe", where the hysteresis loop is lost.  That is, if we set $p=0$ and start the system from the negative consensus state $c^*=0$ and increase $p$, the system follows the lower stable branch until $c^*$ jumps at $p_c$ to a higher value onto the upper stable branch, undergoing a sharp (discontinuous) transition.  Then $c^*$ decreases and approaches $1/2$ as $p \to 1$.  However, if we now decrease $p$ from $c^*(p=1)=1/2$ (reverse path) the system remains in the upper branch ($c^*>1/2$) until it reaches the positive consensus state $c^*=1$ for $p=0$.  Therefore, once the system falls in the upper branch it can never go back to the lower branch and reach a state of negative dominance.  In other words, an initial negative dominance can never be recovered once the system overcomes the threshold $p_c$.  

Figure~\ref{fig:stacionaru_beta1_all_2x2}(b) for $q_{+}=6$ share some similarities with the $q_+=2$ case [Fig.~\ref{fig:stacionaru_beta1_all_2x2}(a)], where the system exhibits no transition for $q_-=1$ and a discontinuous irreversible transition for $q_{-}=5 < q_{+}$. However, for $q_- =8$, the system also shows a reversible hysteresis loop [see Fig.~\ref{fig:double_histeresis}(b) for more detail], in addition to the irreversible hysteresis. Thus, for highly non--linear systems ($q_{\pm} \ge 6$) we observe the existence of a "double hysteresis", one reversible and the other irreversible.  We see that a reversible (symmetric) discontinuous transition already appears for $q_-=q_+=6$, which corresponds to that found in the $q$-VM \cite{Nyczka2012PhaseDriving}. That is why we believe that the reversible hysteresis is probably associated with the classical hysteresis found in the symmetric $q$-VM for $q \ge 6$, whereas the irreversible hysteresis seems to be a consequence of the asymmetry between $+1$ and $-1$ states caused by $q_{+} \neq q_{-}$.  

In Fig.~\ref{fig:double_histeresis} we show in more detail the two types of hysteresis for the set $(q_{+},q_{-})=(6,8)$.  In order to contrast the analytical solution provided by Eq.~(\ref{eq:p_stacionary}) (solid and dashed curves) with MC simulations of the dynamics, we performed simulations in a system of $N=10^6$ agents.  Each symbol corresponds to a single realization of the dynamics starting from a given initial condition, $c(0)=0.0$ (circles) and $c(0)=1.0$ (diamonds).  The dynamics of the system quickly reaches a steady state that depends on the initial condition, where $c(t)$ fluctuates around a stationary value (plateau).  This plateau value was taken as $c^*$, and was estimated as the time average value of $c(t)$ in a time window $\Delta t=1000$ within the stationary state.  In Fig.~\ref{fig:double_histeresis}(a) we see that the numerical values of $c^*$ obtained from simulations (symbols) agree very well with the stable branch (solid curve) that corresponds to a given initial condition. That is, densities $c(t)$ from simulations reach either the lower or the upper branch when they start from $c(0)=0.0$ or $c(0)=1.0$, respectively. The plot displays a double hysteresis: a half hysteresis loop corresponding to the lower branch $0 < p < p_c$ with a discontinuous jump of $c^*$ at the transition point $p_c \simeq 0.055$, and a complete hysteresis loop in the upper branch $0.0805 \lesssim p \lesssim 0.0817$ with discontinuous jumps at the transition points $p_c \simeq 0.0805$ and $0.0817$.  Given that this last hysteresis loop takes place within a very narrow interval of $p$, we zoom in on this range to show the hysteresis in more detail [see  Fig.~\ref{fig:double_histeresis}(b)]. In the bistable regions defined by both hysteresis loops, the system eventually jumps from one to the other stationary state, corresponding to the two stable fixed points.  Far from the transition points, the system remains for long times in the corresponding plateau, but for values of $p$ very close to the transition points, the system remains a short time in each plateau, and thus the time average in a single realization is done over the two different plateaus, leading to a numerical $c^*$ that lies between the two analytical solutions (solid curves).  This can be appreciated in the detailed plot of the reversible hysteresis shown in Fig.~\ref{fig:double_histeresis}(b).

\begin{figure*}[t]
\centering
\includegraphics[width=0.8\textwidth]{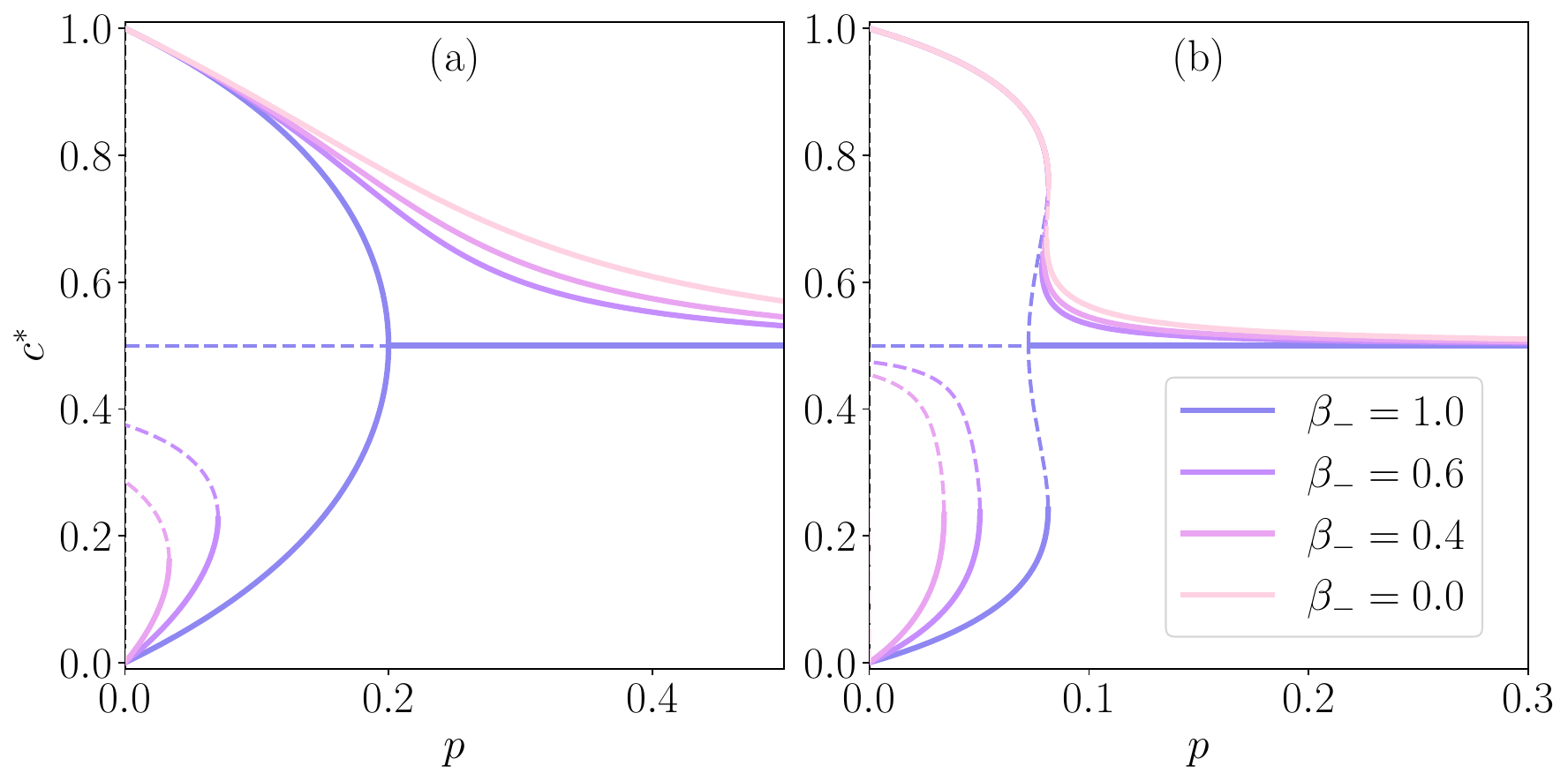}
\caption{\textbf{Extended model}. $c^*$ vs $p$ from Eq.~\eqref{eq:p_stacionary-1} for (a) $q_{+}=q_{-}=2$ and (b) $q_{+}=q_{-}=6$. Each curve corresponds to $\beta_+=1$ and the value of $\beta_{-}$ indicated in the legend.  (a) The continuous transition observed for  $\beta_+=\beta_-=1$ becomes discontinuous for $0<\beta_-<\beta_+$, showing an irreversible hysteresis loop. (b) Besides the discontinuous transition observed for $\beta_+=\beta_-=1$ with an associated reversible loop, an irreversible hysteresis appears for $0<\beta_-<\beta_+$ (double hysteresis).  In both panels, the transitions disappear for $\beta_-=0$.} 
\label{fig:stacionaru_beta_scan}
\end{figure*}

\subsection{Extended model}

We showed in the previous section that by allowing the $q$-VM to have two non-equivalent options $+1$ and $-1$ with different influence group sizes $q_{-}$ and $q_{+} \ne q_{-}$, respectively, breaks the symmetry of the system and leads to discontinuous transitions in the behavior of $c^*$ vs $p$ for values of $q_{-}$ and $q_{+}$ lower than those in the original symmetric model.  Specifically, $(q_{+},q_{-})=(2,3)$ and $(3,2)$ are the sets with the lowest $q$ values that exhibit a discontinuous transition, compared to the value $q_+=q_-=6$ for the symmetric version of the model.  This result leads us to believe that the asymmetry in states has the effect of reducing the minimum value of the group size $q$ required to observe a discontinuous transition.  This idea is consistent with previous findings, where discontinuous transitions were also observed for values of $q$ smaller than $6$ in a $q$-VM with asymmetry in the independence parameter $p$ \cite{Abramiuk-Szurlej2025PairModel}.  In the present model, the asymmetry is not in the independence process but in the influence group.  

Encouraged by these observations, in this section we study an extended version of the model that incorporates an additional asymmetry, expressed in the preference for one of the two options. Instead of adopting a given state every time that the influence group is unanimous in that state, now the change is accepted with state dependent probabilities $\beta_+$ or $\beta_-$. We shall see that this extra source of asymmetry reduces even more the minimum order of non--linearity, given by $q_{+}$ and $q_{-}$, to observe a discontinuous transition. In addition to the "inertia" mechanism introduced by group sizes $q_{+}$ and $q_{-}$ that favors the state with the smallest $q$ value, we add an external field mechanism induced by a difference in the values of $\beta_+$ and $\beta_-$.  When these two mechanisms are at play and favor opposite states, we expect a non-trivial competition between the two, which is worth analyzing.  For example, when $q_{-}>q_{+}$, the largest inertia of $+1$ agents to change state, which tends to lead to a $+1$ dominance, could be compensated by introducing a bias towards $-1$ states, selecting $\beta_{-}>\beta_{+}$.  

To study this phenomenon, we start by writing the rate equation for the evolution of $c$ in the $N \to \infty$ limit, which is similar to Eq.~(\ref{eq:mfa}) but with pre--factors $\beta_{+}$ and $\beta_{-}$ in the gain and loss terms of group interactions, respectively,
\begin{equation}
\begin{split}
\frac{dc}{dt} =\,& (1-p)\left[\beta_{+} (1-c)c^{q_{+}} - \beta_{-}c(1-c)^{q_{-}} \right] + p(1-2c)
\end{split}
\label{eq:mfa-1}
\end{equation}
For the symmetric linear case $q_{\pm}=1$ and $\beta_{+}=\beta_{-}$, the stationary solution is $c^*=1/2$ for $p \ne 1$ and $c^*=c(0)$ for $p=0$.  For the rest of the cases, the stationary solutions are given by the equation
\begin{equation}
    p= \frac{\beta_{+} (1-c^*)c^{*^{q_{+}}}-\beta_{-} c^*(1-c^*)^{q_{-}}}{\beta_{+} (1-c^*)c^{*^{q_{+}}}-\beta_{-} c^*(1-c^*)^{q_{-}}-(1-2c^*)}.
    \label{eq:p_stacionary-1}
\end{equation}
In what follows, we first analyze the effects of having an asymmetry in $\beta_{\pm}$ while keeping the same size of the $q_{\pm}$--panels, and we then study the combined effects of both asymmetries, in the $q_{\pm}$--panel and in $\beta_{\pm}$.    

\begin{figure*}[t]
\centering
\includegraphics[width=0.8\textwidth]{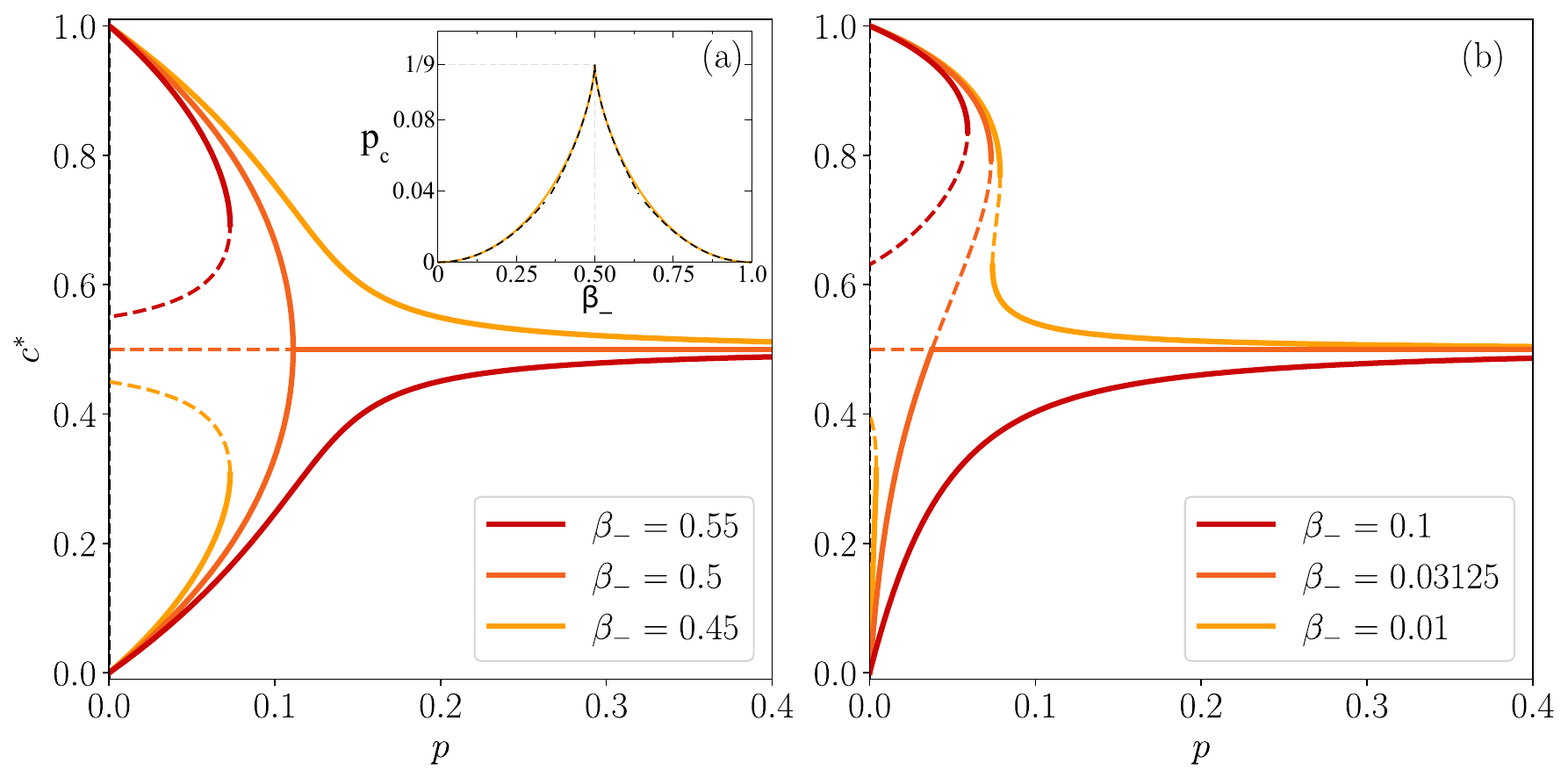}
    \caption{\textbf{Extended model}. $c^*$ vs $p$ from Eq.~(\ref{eq:p_stacionary-1}) for (a) $(q_+,q_-)=(2,1)$ and (b) $(q_+,q_-)=(6,1)$.  Each curve corresponds to $\beta_+=1$ and a different value of $\beta_-$, indicated in the legend. (a) The set $(q_+,q_-)=(2,1)$ corresponds to the lowest order in the $q$-panel that leads to an \emph{irreversible} discontinuous transition at a threshold value $p_c$. The symmetric coexistence solution $c^*=1/2$, stable and unstable, is obtained for $\beta_-=0.5$ from Eq.~(\ref{eq:q_vs_beta}).  The stable solution for $\beta_-=0.5$ (solid line) is the same as that found in the symmetric case $q_+=q_-=2$ and $\beta_+=\beta_-=0.5$ (see main text), and shows a continuous transition. For $\beta_-=0.45$ and $0.55$ the transition becomes discontinuous (lower and upper branches, respectively), characterized by an irreversible hysteresis loop. Inset: Threshold $p_c$ vs $\beta_-$ calculated numerically from Eqs.~\eqref{eq:p_stacionary-2} and \eqref{eq:P-c} (continuous lines). Dashed lines correspond to the analytical approximations Eqs.~\eqref{eq:pc-approx}. The value $p_c=1/9$ of the continuous transition for $\beta_-=0.5$ is indicated by gray lines. (b) The set $(q_+,q_-)=(6,1)$ corresponds to the lowest order in the $q$-panel that leads to a \emph{reversible} discontinuous transition. A double hysteresis is observed for $\beta_-=0.01$, composed by an irreversible hysteresis loop in the lower branch and a reversible hysteresis loop in the upper branch, while the curves for $\beta_-=0.1$ and $0.03125$ show only an irreversible transition (upper branches).}
    \label{fig:demo_of_eq}
\end{figure*}

\subsubsection{$q_{-}=q_{+}$ case}

Before considering the fully asymmetric case, we start by analyzing the simple case $q_{+}=q_{-}$, where the asymmetry comes only from $\beta_+ \ne \beta_-$. The non--linear $q$-VM corresponds to the case $\beta_+=\beta_-=1$.   

The results are shown in Fig.~\ref{fig:stacionaru_beta_scan}, where each panel corresponds to a given value $q_{+}=q_{-}$ and $\beta_{+}=1$, and each curve represents a different value of $\beta_{-}$, as indicated in the legends.  Even though we constructed plots for several values of $q_+=q_-$, we show here the two representative cases $q_{\pm}=2$ [panel (a)] and $q_{\pm}=6$ [panel (b)] that summarize all possible different behaviors found in the plots. Also, the $q_{\pm}=1$ and $\beta_- < \beta_+$ case is peculiar, and corresponds to the linear VM with a bias towards the $+1$ state, leading to a continuous decrease of $c^*$ from $1$ for $p=0$ to $c^*=1/2$ for $p=1$ (not shown).

We see in both panels that the curve for $\beta_-=0$ is larger than $1/2$ for all $p$, reflecting the effect of the bias towards the $+1$ state when $\beta_+>\beta_-$, as expected. In Fig.~\ref{fig:stacionaru_beta_scan}(a) for $q_{\pm}=2$ we observe that the two cases with $0<\beta_-<\beta_+=1$ exhibit a discontinuous transition with an irreversible hysteresis (lower branch), and that the upper branch is above $1/2$ and spans the entire range $p \in [0,1]$, showing the consequence of a positive bias.  Finally, the symmetric continuous transition of the $q_{\pm}=2$ VM is recovered for $\beta_+=\beta_-=1$.
   
In Fig.~\ref{fig:stacionaru_beta_scan}(b) we can see that the irreversible hysteresis loop of the lower branch remains for $q_{\pm}=6$.  This means that a symmetry break in the form of a bias induces a discontinuous transition in the non-linear $q$-VM, which is reminiscent of spin systems subject to an external field. We also observe that an extra discontinuous transition appears due to the high--order of nonlinearity, which corresponds to a reversible hysteresis loop (upper branch), as that observed in the baseline model. The reversible hysteresis was not observed for $q_{\pm}<6$. In other words, a double hysteresis emerges if a bias is introduced in the symmetric $q$-VM, when the level of non--linearity is $q_{\pm}=6$ or higher.

It is important to remark that the bistable region observed in the two panels, where both $\pm 1$ states can dominate, shrinks as $\beta_{-}$ decreases, and vanishes when $\beta_{-}=0$.  For this particular case, the dominance of the $-1$ state (lower branch) disappears, whereas the upper branch representing the $+1$ dominance remains intact.  This means that if the agents of one option are never influenced by a group with the opposite option, the former option dominates for all values of $p$, independently of the initial condition. 

Considering the analysis of the baseline and extended models up to now, we can summarize the results by saying that adding a source of asymmetry in the conformity interaction of the $q$-VM for $q_{\pm} \ge 6$, in the form of different group sizes ($q_+ \ne q_-$) or prestige ($\beta_+ \ne \beta_-$), causes the appearance of a double hysteresis that consists of a reversible (complete loop) hysteresis and an irreversible (half loop) hysteresis.  Besides, the lowest values of the $q_{\pm}$--panels for which the irreversible hysteresis appears are $(q_{+},q_{-})=(2,3)$ and $(3,2)$ in the baseline model ($\beta_{\pm}=1$), and $(q_{+},q_{-})=(2,2)$ in the extended model.  Therefore, the asymmetry in the prestige (bias) gives a lower order of non--linearity to observe a discontinuous transition.

\subsubsection{$q_{-} \ne q_{+}$ case}

Here we analyze the fully asymmetric case $q_{+} \ne q_{-}$ and $\beta_{+} \ne \beta_{-}$. Results are shown in Fig.~\ref{fig:demo_of_eq}. Again, we have chosen two representative sets, $(q_+,q_-)=(2,1)$ [panel (a)] and $(q_+,q_-)=(6,1)$ [panel (b)], with $\beta_+=1$, which exhibit the different behaviors we found by exploring different sets.  Note that, by symmetry in the $\pm 1$ states, the sets $(1,2)$ and $(1,6)$ give the same results as the sets $(2,1)$ and $(6,1)$, respectively. 

Figure~\ref{fig:demo_of_eq}(a) shows the lowest order in the $q_{\pm}$--panel, $(q_+,q_-)=(2,1)$, for which the system exhibits a discontinuous transition. This is observed for $\beta_-=0.45$ and $0.55$, where the respective lower and upper branches correspond to irreversible hysteresis loops.  We also note that the case $\beta_-=0.5$ gives a symmetric continuous transition, as that found in the original $q$-VM with $q=2$.  Indeed, we can check from Eq.~(\ref{eq:mfa-1}) that the asymmetric model with $(q_+,q_-)=(2,1)$ and $\beta_{-}=\beta_{+}/2=\beta/2$ is equivalent to the symmetric model with $q_{+}=q_{-}=2$ and $\beta_{+}=\beta_{-}=\beta/2$, where both lead to the equation
\begin{eqnarray*}
    \frac{dc}{dt} = (1-2c) \left[p-\frac{\beta}{2}(1-p)c(1-c)\right].
\end{eqnarray*}
The stable fixed points are $c^*=\frac{1}{2} \pm \frac{1}{2} \sqrt{\frac{\beta-(8+\beta)p}{\beta(1-p)}}$ for $0 \le p \le p_c$, and $c^*=1/2$ for $p_c \le p \le 1$, where 
\begin{eqnarray}
    p_c \equiv \frac{\beta}{8+\beta} 
    \label{eq:pc-beta}
\end{eqnarray}
is the transition point. The line of fixed points is symmetric around $c^*=1/2$, i.e., $p(c^*)=p(1-c^*)$, as we can also check from Eq.~(\ref{eq:p_stacionary-1}).

This case shows that the asymmetry in the $q_{\pm}$--panel ($q_+=2 q_-$) that favors the $-1$ option is counterbalanced with a positive bias ($\beta_+=2\beta_-$) that favors the $+1$ option, leading to a symmetric solution. Then, one may wonder if there are other combinations of $q_{\pm}$ and $\beta_{\pm}$ for which the perfectly symmetric coexistence state $c^*=1/2$ is a solution. In fact, we can prove that for any $p \ne 1$ the perfectly symmetric density $c^*=1/2$ is a fixed point of Eq.~(\ref{eq:mfa-1}) if the following relation between the parameters is fulfilled: 
\begin{eqnarray}
    \frac{\beta_-}{\beta_+}= 2^{\left(q_- - q_+ \right)}.
    \label{eq:q_vs_beta}    
\end{eqnarray}
Indeed, when we set $c=1/2$ in Eq.~(\ref{eq:mfa-1}), the second term vanishes, and thus when $p \ne 1$ the first term is zero for any combination of the parameters that follow Eq.~(\ref{eq:q_vs_beta}).  For $p=1$, the only fixed point is $c^*=1/2$ for any value of $q_{\pm}$ and $\beta_{\pm}$.  This implies that for a given $q_-<q_+$ that favors the $-1$ state, we can find infinite sets $(\beta_+,\beta_-)$ that counterbalance the preference for the $-1$ state and lead to a symmetric coexistence of both states. However, we note that the fixed point $c^*=1/2$ could be unstable for $p$ small enough, and in this case one or the other state dominates (bistability), as we can see for $\beta_-=0.5$ (dashed line).   

Figure~\ref{fig:demo_of_eq}(b) corresponds to the lowest order in the $q_{\pm}$--panel, $(q_+,q_-)=(6,1)$, for which we observe a double hysteresis.  This is shown for the $\beta_-=0.01$ case where, besides a small irreversible hysteresis loop in the lower branch, a reversible hysteresis loop appears in the upper branch.  This last hysteresis disappears for higher values of $\beta_-$, while the irreversible hysteresis seems to remain for all $\beta_-$ values. We have also added the curve for the value $\beta_-=2^{-5}=0.03125$, obtained from Eq.~(\ref{eq:q_vs_beta}), which leads to the perfectly symmetric solution $c^*=1/2$.

\subsubsection{Analytical results for the $q_{+}=2$ and $q_{-}=1$ case}

The case $(q_{+},q_{-})=(2,1)$ and $\beta_+=1$ is worth examining in some detail because it exhibits a wide range of behaviors while maintaining a low order in the $q_\pm$--panel, which makes some analytical treatment possible. As described in the last section, by keeping $\beta_+=1$ fixed and varying $\beta_-$ we observe an irreversible discontinuous transition for $0<\beta_-<1$, which becomes continuous for $\beta_-=1/2$ [see Fig.~\ref{fig:demo_of_eq}(a)], and disappears for $\beta_-=0$ and $1$. In this case, Eq.~\eqref{eq:p_stacionary-1} becomes
\begin{equation}
    p= \frac{c^*(1-c^*)(c^*-\beta_{-})}{c^*(1-c^*)(c^*-\beta_{-})-1+2c^*}.
    \label{eq:p_stacionary-2}
\end{equation}
We can check from Eq.~\eqref{eq:p_stacionary-2} that $c^*(\beta_-)=1-c^*(1-\beta_-)$, and thus it is enough to consider the case $1/2 \le \beta_- \le 1$. We can also see by inspection that $p=0$ when $c^*$ takes the values $0$, $\beta_-$ and $1$, and that $p=1$ when $c^*=1/2$. These points determine the lower branch $c^* \in [0,1/2]$ and the upper branch $c^* \in [\beta_-,1]$, where $p$ takes physical values $0\le p \le1$ [see curves for $\beta_-=0.55$ in Fig.~\ref{fig:demo_of_eq}(a)]. The nonphysical values $p>1$ and $p<0$ are obtained for $c^* \in (1/2,\Tilde{c})$ and $c^* \in (\Tilde{c},\beta_-)$, respectively, where $\Tilde{c}$ is the real root of the denominator of Eq.~\eqref{eq:p_stacionary-2}. In the lower branch, $c^*$ increases smoothly from $c^*(p=0)=0$ to $c^*(p=1)=1/2$. In contrast, the upper branch shows a discontinuous transition at a value $p_c$ that corresponds to the maximum of $p(c^*)$ at $c^*_c$, i.e. $p(c^*_c)=p_c$, Thus, $c^*_c$ must satisfy the condition $\frac{\partial p}{\partial c^*}|_{c^*_c}=0$, which leads to the following equation for $c^*_c$:
\begin{eqnarray}
    (c^*_c-\beta_-)(1-2c^*_c)^2+(1-2\beta_-)c^*_c(1-c^*_c)=0.
    \label{eq:P-c}
\end{eqnarray}
The three roots of the cubic polynomial in Eq.~\eqref{eq:P-c} can be found explicitly, but we omit the expressions here because they are rather cumbersome. The real root corresponds to the line of transition points $c^*_c(\beta_-)$ for $0 \le \beta_- \le 1$. By replacing this real expression for $c^*_c(\beta_-)$ into Eq.~\eqref{eq:p_stacionary-2} we obtain a closed expression for the transition points $p_c(\beta_-)$, which are plotted in the inset of Fig.~\ref{fig:demo_of_eq}(a) with a solid line. 

In the following, we derive rather simple —although approximate— expressions for $p_c$ as a function of $\beta_-$, for $\beta_-$ in the vecinity of $0$, $1/2$ and $1$. We observe from Eq.~\eqref{eq:P-c} that for $\beta_-=1$ the real root is $c^*_c=1$, meaning that the discontinuous transition disappears in the $\beta_- \to 1$ limit, as expected.  We also see that $c^*_c=1/2$ when $\beta_-=1/2$, which corresponds to the transition point $p_c=1/9$ of the continuous transition for the case $\beta_-=\beta_+/2=\beta/2=1/2$ [Eq.~\eqref{eq:pc-beta}], obtained by taking the limit $c^* \to 1/2$ in Eq.~\eqref{eq:p_stacionary-2}. 

Given $c^* \in [\beta_-,1]$ in the upper branch, when $\beta_- \lesssim 1$ we can write $c^*_c=1-\epsilon$, with $0< \epsilon \ll 1$. Taylor expanding Eq.~\eqref{eq:P-c} up to first order in $\epsilon$ and solving for $\epsilon$, we obtain $\epsilon=\frac{1}{2}(1-\beta_-)/(2-\beta_-)$, and thus
\begin{eqnarray}
    c^*_c \simeq \frac{(3-\beta_-)}{2(2-\beta_-)} ~~~ \text{for} ~~~ \beta_- \lesssim 1.
    \label{eq:c-approx-1}
\end{eqnarray}
Similarly, we write $c^*_c=1/2+\epsilon$ when $\beta_- \gtrsim 1/2$, and obtain from Eq.~\eqref{eq:P-c} the equation $(1-2\beta_-)(1+4\epsilon^2)+16 \epsilon^3=0$.  For fixed $\epsilon >0$ and taking $\beta_-$ very close to $1/2$, the term of order $\epsilon^2$ becomes negligible compared to the other two terms; thus, we get $\epsilon \simeq [(2\beta_--1)/16]^{1/3}$. Then,
\begin{eqnarray}
    c^*_c \simeq \frac{1}{2} + \left( \frac{2\beta_--1}{16} \right)^{1/3} ~~~ \text{for} ~~~ \beta_- \gtrsim 1/2.
    \label{eq:c-approx-2}
\end{eqnarray}
Finally, plugging the approximate expressions for $c^*_c$ from Eqs.~\eqref{eq:c-approx-1} and \eqref{eq:c-approx-2} into Eq.~\eqref{eq:p_stacionary-2} and expanding, respectively, up to second order in $(1-\beta_-)$ and up to second order in $(2\beta_{-}-1)^{2/3}$ we arrive at
\begin{eqnarray}
    \label{eq:pc-approx}
    p_c &\simeq& \tfrac{1}{8} (3-\beta_-)(1-\beta_-)^2 ~~~ \text{for } \beta_- \lesssim 1, \nonumber \\
    p_c &\simeq& \tfrac{1}{8} (2+\beta_-)\beta_-^2 ~~~ \text{for } \beta_- \gtrsim 0, \\
    p_c &\simeq& \tfrac{1}{9} - \tfrac{2^{7/3}}{27} |2\beta_{-}-1|^{2/3} 
     + \tfrac{2^{5/3}}{81} |2\beta_{-}-1|^{4/3} ~ \text{for } \beta_- \simeq 1/2. \nonumber
\end{eqnarray}
Approximations from Eqs.~\eqref{eq:pc-approx} are shown by dashed lines in the inset of Fig.~\ref{fig:demo_of_eq}(a).

\section{Summary and Conclusions}

We introduced and studied a modified version of the $q$-VM with independence, in which the influence group size ($q_s \ge 1$ integer) depends on the state (opinion) of an agent, denoted as $q_+$ and $q_-$ for positive and negative opinions, respectively. This formulation aims to capture the negativity bias phenomenon, wherein a smaller number of negative influences ($q_- < q_+$) is sufficient to induce a shift from a positive to a negative opinion. We further extended the model by incorporating two preference parameters, $\beta_+$ and $\beta_-$, which modulate the probability of adopting positive and negative opinions, respectively. Both models were investigated within a mean-field framework and validated through numerical simulations. The introduction of asymmetry between opinions via these parameters gives rise to substantially richer dynamical behavior, as compared to the original symmetric $q$-VM, as we describe below.

We characterized the system at the collective level by analyzing the fraction of positive opinions in the stationary state, denoted as $c^*$, as a function of the independence parameter $p$. Our results reveal several types of transitions. Continuous transitions occur for $1 \leq q_- \leq 5$ and $q_+ \geq 2$; these transitions are symmetric with respect to $c^* = 1/2$ when $2 \leq q_+ = q_- \leq 5$ and for $(q_+, q_-) = (2, 1)$, and asymmetric otherwise. Discontinuous irreversible transitions arise for $q_+ \geq 2$ and $q_- \geq 1$, whereas discontinuous reversible transitions are observed when $q_+ \geq 6$ and $q_- \geq 1$. Finally, the system exhibits no transition for $q_- = 1$, $q_+ \geq 1$, $\beta_+ = 1$, and $\beta_- = 0, 1$. It is important to note that, because the system is invariant under the transformation $+ \leftrightarrow -$, identical behavior is obtained when interchanging the values of $q_+$ and $q_-$, and $\beta_+$ and $\beta_-$.

Although each type of transition is associated to a specific set of parameter values, two main behaviors emerge depending on the system’s nonlinearity determined by $q_{\pm}$. First, when at least one influence group has size $q_{\pm} = 2$ or larger, the system exhibits an \textit{irreversible discontinuous transition} at a threshold value $p_c$. This irreversibility is characterized by a half hysteresis loop leading to a \textit{cusp catastrophe}: as $p$ increases and overcomes $p_c$, the system departs from one stable branch and settles on another, remaining trapped and unable to return to the original $c^{*}$ values. Consequently, once the independence threshold $p_{c}$ is surpassed, the system cannot recover its initial state. This mechanism may help explain phenomena such as the irreversible damage of a product’s reputation in online markets. Second, when at least one influence group is relatively large ($q_{\pm} = 6$ or greater), the system exhibits a \textit{reversible discontinuous transition}, reminiscent of the original $q$-VM and associated with a complete hysteresis loop. Interestingly, within the range $q_{\pm} \ge 6$ and $q_{\mp} \ge 1$, the system can display both types of discontinuous transitions simultaneously, one reversible and one irreversible, resulting in a \textit{double hysteresis} behavior.

A similar phenomenon, that is, irreversible hysteresis for $q \geq 2$, was reported recently in the biased-independence $q$-voter model \cite{Abramiuk-Szurlej2025PairModel}, although the source of asymmetry in that model was different. In that study, the size of the influence group $q$ was fixed, but during independent behavior the positive opinion (interpreted as the adopted state) was favored. This demonstrates a general tendency: whenever the symmetry between opinions is broken, regardless of how this asymmetry is introduced, the collective dynamics can become path-dependent and exhibit irreversible hysteresis. What is, however, very interesting here, both from the modeling and social point of view, is that in the present opinion-dependent $q$-voter model the asymmetry is embedded in the interaction rule itself, through opinion-dependent influence group sizes. This captures the essence of the well-documented negativity bias, where negative information or experiences exert a stronger impact than positive ones. In our model this is reflected by the fact that a small group of dissenters may be enough to overturn a consensus, whereas rebuilding agreement requires much stronger support. The resulting irreversible or double hysteresis can thus be interpreted as a formal representation of lasting reputational damage or asymmetric recovery after crises, highlighting how simple local asymmetries in social influence can lead to complex and persistent collective effects.

\begin{acknowledgments}
This research was partially funded by the National Science Centre, Poland Grant 2019/35/B/HS6/02530 (MD and KSW).

\end{acknowledgments}
\section*{Author Declarations}
The authors have no conflicts to disclose.

\section*{Data Availability Statement}
The source code in Julia, which allows for the reproduction of all results in this publication, is publicly available at 
\href{https://github.com/TheMik1999/Negativity_Bias.git}{GitHub}.

\bibliography{bibliography}
\end{document}